# Scalar - vector soliton fiber lasers

Zhichao Wu,[1,2,+] Deming Liu,[1,+] Lei Li,[2] Yiyang Luo,[1] Dingyuan Tang,[2] Deyuan Shen,[2] Ming Tang,[1] Songnian Fu,[1,*] Luming Zhao[2,*]

[1]National Engineering Laboratory of Next Generation Internet Access System, School of Optical and Electronic Information, Huazhong University of Science and Technology, Wuhan, Hubei, 430074, China

[2] Jiangsu Key Laboratory of Advanced Laser Materials and Devices, School of Physics and Electronic Engineering, Jiangsu Normal University, Xuzhou, Jiangsu, 221116, China

+These authors contributed equally to this work and should be considered co-first authors

*Correspondence: *songnian@hust.edu.cn*, *zhaoluming@jsnu.edu.cn*

**ABSTRACT:** Rapid progress in passively mode-locked fiber lasers is currently driven by the recent discovery of vector feature of mode-locking pulses, namely, the group velocity-locked vector solitons, the phase locked vector solitons, and the high-order vector solitons. Those vector solitons are fundamentally different from the previously known scalar solitons. Here, we report a fiber laser where the mode-locked pulse evolves as a vector soliton in the strong birefringent segment and is transformed into a regular scalar soliton after the polarizer within the laser cavity. The existence of solutions in a polarization-dependent cavity comprising a periodic combination of two distinct nonlinear waves is novel and likely to be applicable to various other nonlinear systems. For very large local birefringence, our laser approaches the working regime of vector soliton lasers, while it approaches scalar soliton fiber lasers under the conditions of very small birefringence.

## INTRODUCTION

Passively mode-locked fiber laser as a flexible source of ultrafast optical pulses has been widely investigated over past decades, due to its advantages of compact size, high stability and simple structure[1-3]. In addition, such lasers act as a convenient experimental platform for the investigation of nonlinear waves subject to periodic boundary conditions and easy operation. In order to achieve self-started mode-locking, various techniques have been proposed, including the nonlinear polarization rotation (NPR)[4], nonlinear loop mirror (NOLM)[5], semiconductor saturable absorber mirror (SESAM)[6], and various two-dimensional optical materials[7-9]. Among those mode-locked techniques, the NPR technique has been commonly adopted, due to its comprehensive phenomena and easy implementation[10].

Because of the requirement of the polarization dependent component in a NPR cavity, theoretically a scalar model is enough to describe the fiber laser even it is well known that a single mode fiber (SMF) always supports two orthogonal polarization modes. The theoretical results from the scalar model agree with the experimental observations very well. Therefore, the solitons generated from a NPR cavity are always considered as scalar solitons[11]. Light polarization is a traditional but newly discovered very useful feature for telecommunications[12,13]. Polarization properties of ultrashort pulses are newly explored, which shows a distinctive regime for pulse generation. Therefore, a strong motivation to search for this new regime arises.

On most occasions, due to the asymmetrical structure and bending, the birefringence of SMF gives rise to the differences of two orthogonal polarization modes in term of both phase and group velocities[14]. On the other hand, when the soliton propagates periodically within a laser cavity, phase locking between two orthogonal polarization modes can be achieved, leading to the generation of vector soliton[15-17]. Consequently, for a fiber laser without polarization dependent components, it favours the generation of vector solitons. The dependence of the cavity birefringence versus the vector soliton generation and features are comprehensively investigated at different wavelengths[18,19].

In order to generate vector solitons, polarization dependence of the laser cavity should be avoided as much as possible. In the previous researches, SESAM and carbon nanotube (CNT) are the most widely used mode-lockers to obtain vector soliton in the passively mode-locked fiber lasers[20-25]. Topological insulators can also be used as the saturable absorber for vector soliton generation. Because of the absence of the polarization constraint, two orthogonal polarization modes in the fiber laser can develop freely. Therefore, much enriched characteristics emerge from vector soliton fiber laser compared with the scalar one[26-27].

Local strong birefringence does not affect the polarization dependence. Therefore, it is possible to introduce fast polarization evolution during pulse propagation in a fiber laser with low average birefringence. Visible time delay may accumulate if a local strong birefringence is applied. Different from the mode-locking using physical saturable absorbers, for the case of NPR-based mode-locking, the use of inline polarizer restrains the pulse polarizations. Therefore, it is always considered that such type of mode-locked lasers could only generate scalar solitons. In fact, the pulse can exhibit vector characteristics at specific positions within the laser cavity. Here, we present our theoretical prediction and experimental demonstration, which shows an entirely new operation regime where the pulse develops into vector along the local strong birefringence fiber segment, and following polarization filtering, evolves into a scalar soliton and maintains its polarization in the rest of the cavity, when the fiber has weak birefringence. The scalar and vector characteristics are verified from the variation of corresponding optical spectra and autocorrelation traces by the use of an all-fiber polarization resolved measurement. All mode-locked lasers to date are considered to have a single type of polarization during pulse propagation within the cavity. However, for our laser, distinctly different vector solitons and scalar solitons co-exist, indicating that transitions between them are possible. The experimental results agree well with our numerical simulation. Remarkably, this configuration is extremely robust against perturbations. Although the pulse experiences nonlinear effects strong enough to cause unprecedented, kink-jumped variations of the polarization, the laser shows excellent short- and long-term stability.

**MATERIALS AND METHODS**

The numerical simulation is carried out based on the fiber laser specially designed as shown in Fig. 1. A 2.2-m erbium-doped fiber (EDF) with a group velocity dispersion (GVD) parameter of -18 (ps/nm)/km is utilized as the gain medium, which is pumped via 980/1550 wavelength division multiplexing (WDM) by a 976-nm laser diode (LD) with maximum output power of 750mW. The fiber pigtails of optical components are 15-m standard SMF with a GVD parameter of 17 (ps/nm)/km. Therefore, the ring cavity is in anomalous dispersion region, and the total cavity length is around 17.2m. The beat length of the EDF is about 5cm and that of the standard SMF is about 10m. The absorption coefficient of the EDF is around 21.9dB/m at 980nm. Two polarization controllers (PCs) together with an inline polarizer are used to implement the NPR-based mode-locking. The polarization-independent isolator is used to guarantee unidirectional propagation and suppress detrimental reflections.

For the ease of simultaneously monitoring both the scalar and vector soliton within the same laser cavity, two optical couplers (OCs) with a power ratio of 20:80 are separately set before and after the polarizer. Following the output, PC3 along with an inline polarization beam splitter (PBS) is used to implement all-fiber polarization resolved measurement. Finally, an optical spectrum analyzer (OSA, Yokogawa AQ6370C-20) with a resolution of 0.02nm and a 1-GHz oscilloscope (OSC, Agilent DSO9104A) together with a 2-GHz photodetector (PD) are applied to monitor the optical spectrum and the temporal trace, respectively. Additionally, the pulse profile can be measured by a commercial autocorrelator (Femtochrome FR-103HS).

The laser operation is simulated based on the coupled Ginzburg-Landau equations, which describe the pulse propagation. The pulse propagation in fibers is governed by:

$$\begin{cases} \dfrac{\partial u}{\partial z} = i\beta u - \delta \dfrac{\partial u}{\partial t} - \dfrac{ik''}{2}\dfrac{\partial^2 u}{\partial t^2} + \dfrac{ik'''}{6}\dfrac{\partial^3 u}{\partial t^3} + i\gamma(|u|^2 + \dfrac{2}{3}|v|^2)u + \dfrac{i\gamma}{3}v^2 u^* + \dfrac{g}{2}u + \dfrac{g}{2\Omega_g^2}\dfrac{\partial^2 u}{\partial t^2} \\ \dfrac{\partial v}{\partial z} = -i\beta v + \delta \dfrac{\partial v}{\partial t} - \dfrac{ik''}{2}\dfrac{\partial^2 v}{\partial t^2} + \dfrac{ik'''}{6}\dfrac{\partial^3 v}{\partial t^3} + i\gamma(|v|^2 + \dfrac{2}{3}|u|^2)v + \dfrac{i\gamma}{3}u^2 v^* + \dfrac{g}{2}v + \dfrac{g}{2\Omega_g^2}\dfrac{\partial^2 v}{\partial t^2} \end{cases}$$

where $u$ and $v$ are normalized envelopes of the pulse along two orthogonal polarizations. $2\beta = 2\pi\Delta n/\lambda$ is the wave number difference between two modes. $2\delta = 2\beta\lambda/2\pi c$ is the inverse group velocity difference. $k'$ is the second-order dispersion coefficient, $k''$ is the third-order dispersion coefficient, and $\gamma$ is the fiber nonlinearity coefficient. $g$ and $\Omega_g$ represent the saturable gain coefficient and gain bandwidth of the EDF, respectively. We consider the gain saturation as:

$$g = G \exp\left[-\frac{\int(|u|^2 + |v|^2)dt}{P_{sat}}\right]$$

where $G$ is the small signal gain coefficient and $P_{sat}$ is the normalized saturation energy.

The parameters are set as follows:

$\Omega_g = 20 nm$, $\gamma = 3W^{-1}km^{-1}$, $k'' = -23 ps^2/km$, $k''' = -0.13 ps^3/km$,

$L = 17m$, $L_b^{SMF} = 8.5m$, $L_b^{EDF} = 5cm$

## RESULTS

**a. Numerical Simulations**

The mode-locking can be achieved under the condition of G=1300. It is found that at the OC1, steady pulses with linear polarization are obtained. In other words, no matter how much the phase delay is introduced by the PC3, the horizontal spectrum always resembles the vertical spectrum after the all-fiber polarization resolved measurement, as shown in Fig. 2. However, it is found that at the OC2, two polarization-resolved outputs after the PBS have different spectra and pulse profiles, as shown in Fig. 3. A pulse with a two-humped peak along one polarization and a single-humped along the orthogonal polarization can be observed, as shown in Fig. 3(a), Fig. 3(c), and Fig. 3(e). The spectral dip is formed due to the interference between two humps. Fig. 3(b) shows a strong spectral dip at central wavelength, indicating a 180-degree phase difference. In addition, duo to the gradually change made to the phase delay introduced by the PC3, the dip moves away from the central wavelength, as shown in Fig. 3(d) and Fig. 3(f). The details from Fig. 3 suggest that the pulse obtained at the OC2 is a vector soliton where the two orthogonal components of the vector soliton have certain time delay[28].

Obviously, the inline polarizer acts as a polarization selection component in the cavity. Therefore, the pulse after the polarizer will always be shaped into a scalar soliton, and then recirculates within the cavity. A linearly polarized pulse or a scalar soliton will be transformed into an elliptically polarized pulse then back to its originally linear polarization at every beat length. The time delay between two orthogonal polarizations arising from the fiber birefringence is generally neglected due to the weak birefringence. However, it is not the case here. Due to the strong local birefringence of the EDF, the time delay accumulating for two polarizations during propagation could not be ignored any more. Although the central wavelength of two orthogonal polarizations still keeps the same, the evident time delay makes them form a vector soliton. The time delay will be extinguished when the vector soliton goes through the polarizer, when one polarization is vertical to the axis of the polarizer. Consequently, a cycle is fulfilled. Therefore, the output from OC2 shows features distinct from that of OC1, after the all-fiber polarization resolved measurement. After passing through the PBS, under a specific condition, the projection of the vector soliton could have single peak with maximum intensity while the orthogonal projection shows a structure of two humps with a 180-degree phase difference, leading to the evident spectral dip at the central wavelength, as shown in Fig. 3(b).

**b. Experimental Results**

Encouraged by the simulations, the laser of Fig. 1 is built. We set the pump power at 160mW, and carefully adjust the PC1 and PC2. The stable mode-locking can be easily achieved. To exclude the complications caused by soliton interactions, we reduce the number of solitons in the cavity by carefully decreasing pump power to 100mW, so that only one soliton remains in cavity. The averaged output power from OC1 and OC2 are 1.8mW and 2.6mW, respectively. The reduced output power comes from the loss caused by the polarizer.

Experimentally we reproduce all the details from the simulations. The direct output from the OC1, two selected polarization-resolved outputs are shown in Fig. 4. The optical spectra of two resolved polarizations after the PBS always move simultaneously with one rising up and the other falling down. We can achieve the maximum power of one direction, while the other is reduced to the lowest, as shown by the red and blue solid lines in Fig. 4. It is obvious that the spectra only have intensity difference while the spectral profile is maintained. The central wavelengths of the direct output and of the polarization-resolved outputs are the same. The polarization extinction ratio (ER) of PBS is about 30dB. Therefore, it is reliable to conclude that the direct output from the OC1 has fixed linear polarization, which can be defined as the scalar soliton. Meanwhile, Fig. 5 shows the corresponding autocorrelation traces and pulse-trains. The full-widths at half maximum (FWHM) of two orthogonal polarizations are both 0.9ps if a $sech^2$ pulse profile is assumed. The pulse-trains, as shown in the inset, present consistent peak intensity with a repetition rate of 11.76 MHz, which corresponds to the cavity length. The intensity of the pulse-trains of the two resolved polarizations simultaneously rises and decreases when the PC3 is tuning. When one resolved polarization reaches the maximum, the orthogonal polarization is close to zero.

The output from OC2 is as-well examined following the same procedure. By adjusting PC3, we can no longer observe completed extinction in any polarization direction. Instead, when the spectrum of one direction decline to the lowest, it appears an evident spectral dip at the central wavelength, as shown by blue line in Fig. 6. The red line illustrates that no such dip appears at the spectrum of its orthogonal direction. Additionally, the dip could be moved to either edges when PC3 is rotated, as shown by the green and purple dot line. To identify the source of the spectral dip, we further measure the corresponding autocorrelation traces, as shown in Fig. 7. Note that the vertical direction has a double-humped intensity profile, which is similar to a bound state soliton[29-31]. The FWHM of the humps is about 0.48ps, and the separation between the humps is about 0.64ps. The horizontal direction has a sech$^2$ profile with a FWHM of 0.44ps. Based on the autocorrelation measurement, we can infer that the direct output from the OC2 is a vector soliton with two orthogonal components having 0.64ps time delay[28]. The spectral dip is formed due to the anti-phase projections of the vector soliton along the two polarization directions of the PBS. The deepest dip at the central wavelength indicates that the two anti-phase projections have the same amplitude.

**CONCLUSIONS**

In conclusion, we propose a simple and convenient configuration to investigate a novel mode-locking regime of an erbium-doped fiber laser, with scalar soliton and vector soliton propagation occurring in each half of the cavity. The vector solitons are the first experimental observation inside a fiber laser using NPR technique. It is caused by the strong local birefringence. In fact, we check the scalar-vector soliton fiber laser operation with another scheme: first we use all fibers of weak birefringence and find the pulse features from the OC1 and the OC2 are the same and only the scalar solitons are obtained from either output; then we inserted a short segment (~10 cm) polarization-maintaining fiber between the polarizer and the OC1 and the vector solitons are obtained both from the OC1 and the OC2. The combination of a fiber segment with strong birefringence to undo the polarization selection from the polarizer is the key step for the experimental demonstration of a vector soliton from a NPR fiber laser. The transitions between the scalar and vector solitons are inherently interesting due to their vastly different characteristics. Limited by increasing local birefringence, the laser becomes identical to a vector soliton laser mode locked by polarization-independent saturable absorber. In the other extreme of vanishing fiber section of strong birefringence, the cavity becomes identical to that of a scalar soliton laser. Thus, this new mode-locking regime sits at a nexus of all other known regimes of operation. Scalar solitons and vector solitons can coexist in a fiber laser, which greatly enriches the understanding of soliton dynamics.

**Acknowledgements**

This research has been carried out at Jiangsu Normal University, which is supported by National Key Scientific Instrument and Equipment Development Project (2013YQ16048702), National Natural Science Foundation of China (61275109, 61275069, 61331010).

**Author contributions**

Z. W. conceived the study, performed experiments, analyzed the data and co-wrote the paper. D. L. analyzed the data and co-wrote the paper. L. L. performed numerical simulations, contributed to the discussion of the results and paper writing. Y. L. contributed to the discussion, interpretation of the data. D. T. contributed to the data analysis, discussion and paper writing. D. S. contributed to the data analysis, discussion and paper writing. M. T. contributed to the data analysis, discussion and paper writing. S. F. contributed to data analysis, interpretation of data and paper writing. L. Z. conceived the experiment and numerical simulations, contributed to the discussion of the results and paper writing.

**Figures**

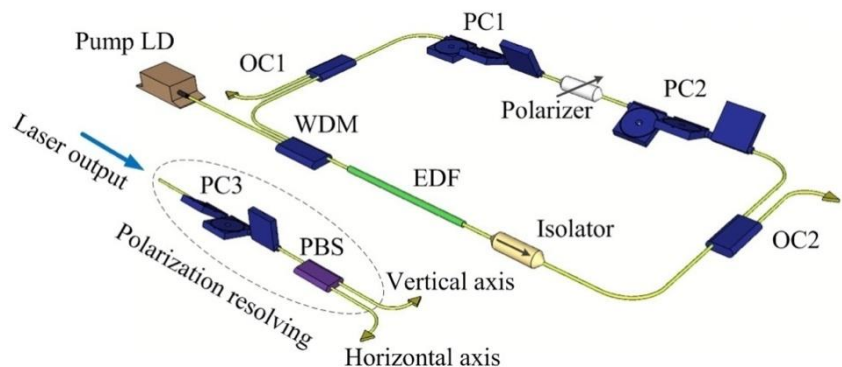

Fig.1. Schematic illustration of the experimental setup: the laser cavity and polarization resolved measurement. LD, laser diode. PC, polarization controller. OC, optical coupler. WDM, wavelength division multiplexer. EDF, erbium-doped fiber. PBS, polarization beam splitter.

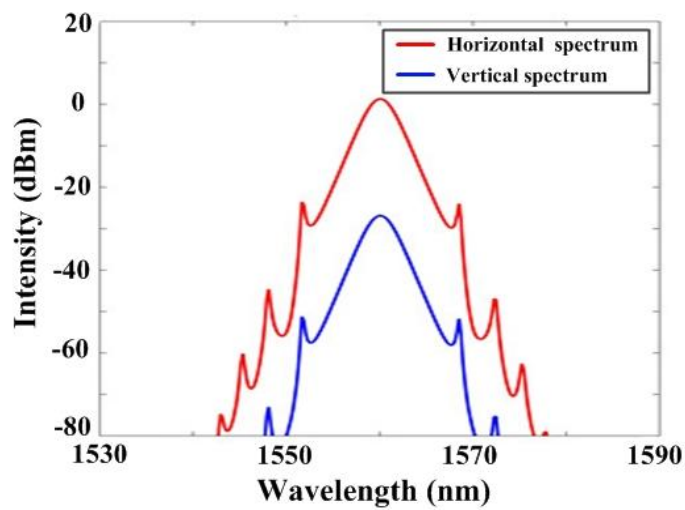

Fig.2. Numerical simulation of the output spectrum of the polarization-resolved projections from OC1. Horizontal component (red line) and vertical spectrum (blue line).

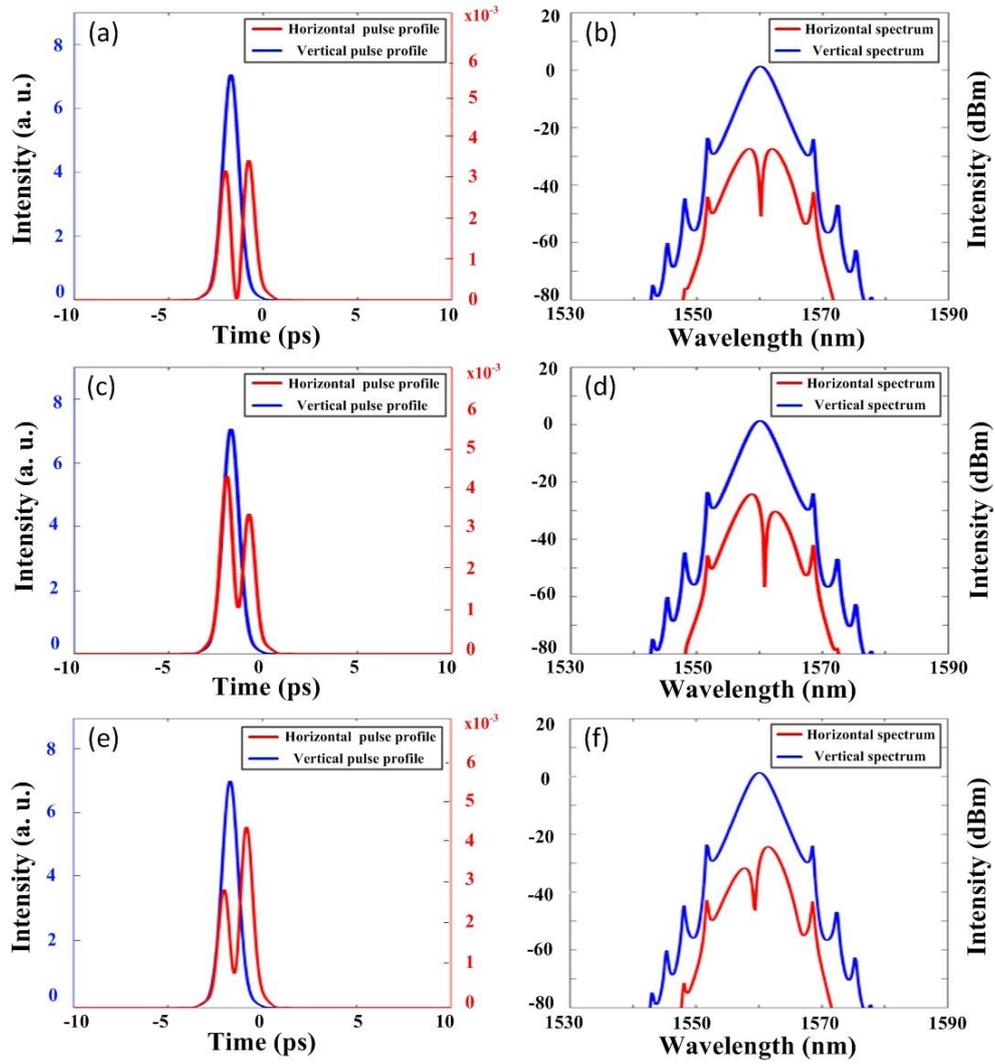

Fig.3. Numerical simulation of the polarization-resolved projections from OC2. (a)(c)(e) intensity profiles and (b)(d)(f) corresponding optical spectra. (a)(b) Two orthogonal components with a 180-degree phase difference. (c)(d)(e)(f) The movement of pulse profile and spectral dip by gradually changing the phase delay introduced by the PC3.

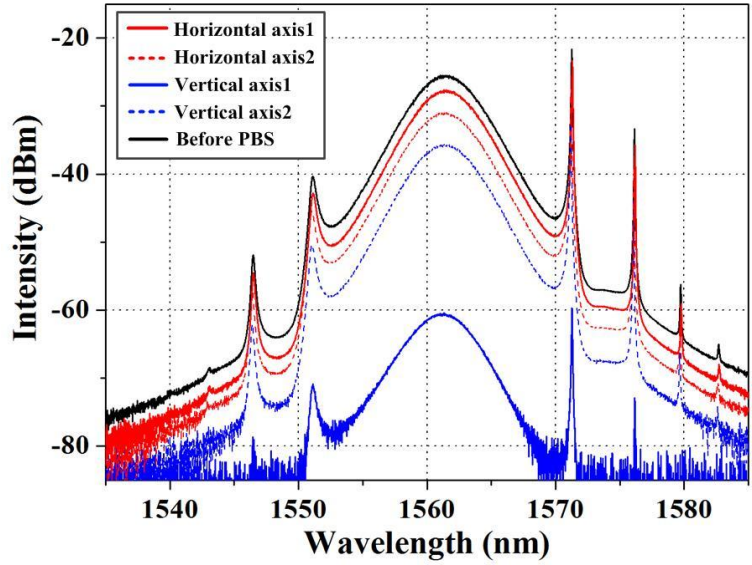

Fig.4. Experimentally observed optical spectra of the two orthogonal axes from OC1 before and after the PBS. Exampled normal state: the horizontal axis (red dot) and vertical axis (blue dot). Maximum-minimum state: the horizontal axis (red solid), vertical axis (blue solid) and total spectrum (black)

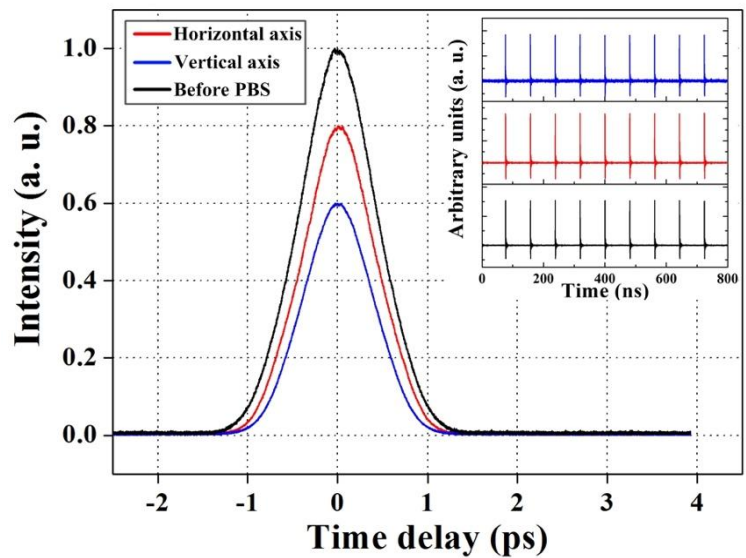

Fig. 5.Experimental autocorrelation traces and pulse-trains from OC1 before and after PBS. Horizontal axis (red), vertical axis (blue) and total pulse profile (black).

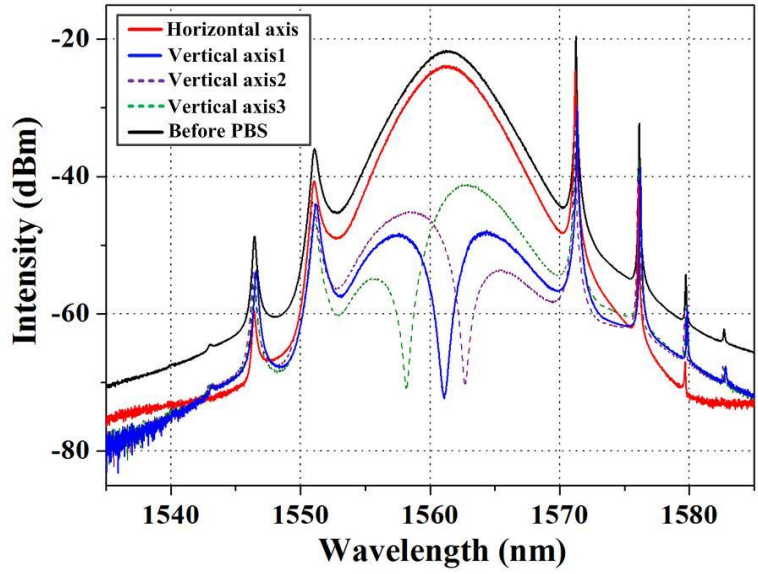

Fig. 6. Experimental optical spectra of the two orthogonal axes from OC2 before and after the PBS. The horizontal axis (red), vertical axis when two polarized components have a 180-degree phase difference (blue), vertical axis with dip movement (purple and green) and total spectrum (black).

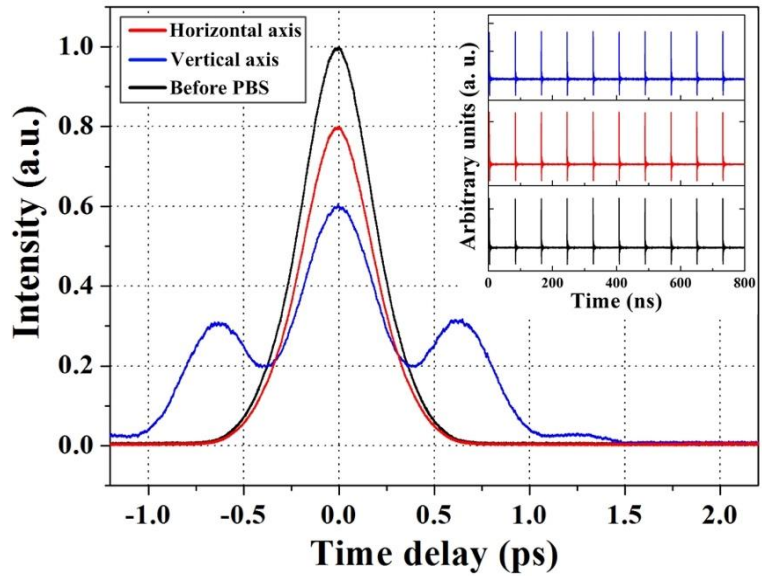

Fig. 7. Experimental autocorrelation traces and pulse-trains from OC2 before and after PBS. Horizontal axis (red), vertical axis (blue), and total pulse profile (black).